\documentclass[fleqn,usenatbib]{mnras}

\usepackage{newtxtext,newtxmath}

\usepackage[T1]{fontenc}

\DeclareRobustCommand{\VAN}[3]{#2}
\let\VANthebibliography\thebibliography
\def\thebibliography{\DeclareRobustCommand{\VAN}[3]{##3}\VANthebibliography}

\usepackage{graphicx}	
\usepackage{amsmath}	

\newcommand{\Msunyr}{M$_{\sun}$yr$^{-1}$}

\newcommand{\Msun}{M$_{\sun}$}
\newcommand{\Mdot}{\dot{M}}
\newcommand{\Mdotss}{\dot{M}_{\mathrm {ss}}}

\title[The distribution of protoplanetary disc accretion rates]{The distribution of accretion rates as a diagnostic of protoplanetary disc evolution}

\author[Alexander et al.]{Richard Alexander,$^{1}$\thanks{E-mail: richard.alexander@leicester.ac.uk}
Giovanni Rosotti,$^{1,2,3}$
Philip J.\,Armitage,$^{4,5}$
Gregory J.\,Herczeg,$^{6,7}$\newauthor
Carlo F.\,Manara$^{8}$ and
Beno\^it Tabone$^{9}$
\\
$^{1}$School of Physics \& Astronomy, University of Leicester, University Road, Leicester, LE1 7RH, UK\\
$^{2}$Dipartimento di Fisica ``Aldo Pontremoli'', Universit\`a degli Studi di Milano, via G.Celoria 16, I-20133 Milano, Italy\\
$^{3}$Leiden Observatory, Leiden University, P.O.\,Box 9513, 2300 RA Leiden, the Netherlands\\
$^{4}$Center for Computational Astrophysics, Flatiron Institute, 162 Fifth Avenue, New York, NY 10010, USA\\
$^{5}$Department of Physics and Astronomy, Stony Brook University, Stony Brook, NY 11794, USA\\
$^{6}$Kavli Institute for Astronomy and Astrophysics, Peking University, No.5 Yiheyuan Road, Haidian District, Beijing 100871, People's Republic of China\\
$^{7}$Department of Astronomy, Peking University, No.5 Yiheyuan Road, Haidian District, Beijing 100871, People's Republic of China\\
$^{8}$European Southern Observatory, Karl-Schwarzschild-Strasse 2, 85748, Garching bei M{\"u}nchen, Germany\\
$^{9}$Universit\'e Paris-Saclay, CNRS, Institut d’Astrophysique Spatiale, 91405 Orsay, France
}

\date{Accepted 2023 June 27. Received 2023 June 27; in original form 2023 January 24}

\pubyear{2023}

\begin{document}
\label{firstpage}
\pagerange{\pageref{firstpage}--\pageref{lastpage}}
\maketitle

\begin{abstract}
We show that the distribution of observed accretion rates is a powerful diagnostic of protoplanetary disc physics. Accretion due to turbulent (``viscous'') transport of angular momentum results in a fundamentally different distribution of accretion rates than accretion driven by magnetised disc winds. We find that a homogeneous sample of $\gtrsim$300 observed accretion rates would be sufficient to distinguish between these two mechanisms of disc accretion at high confidence, even for pessimistic assumptions. Current samples of T Tauri star accretion rates are not this large, and also suffer from significant inhomogeneity, so both viscous and wind-driven models are broadly consistent with the existing observations. If accretion is viscous, the observed accretion rates require low rates of disc photoevaporation ($\lesssim$\,$10^{-9}$\,\Msunyr). Uniform, homogeneous surveys of stellar accretion rates can therefore provide a clear answer to the long-standing question of how protoplanetary discs accrete. 
\end{abstract}




\section{Introduction}
Planets form in cold discs of dust and gas around newly-formed stars. These discs dominate the mass and angular momentum of forming planetary systems, as well as providing the raw material for planets. Understanding protoplanetary disc evolution is therefore a critical ingredient of any predictive theory of planet formation. 

The long-standing paradigm is that protoplanetary disc accretion is due to turbulent transport of angular momentum, driven by the magnetorotational instability (MRI) \citep{bh91,balbus11}. The picture of protoplanetary discs as ``viscous'' accretion discs is well-established \citep[e.g.,][]{lbp74,hartmann98}, and accretion disc theory can plausibly explain many observed properties of protoplanetary discs \citep[e.g.,][]{wc11}. The efficiency of turbulent transport is parametrized in terms of the \citet{ss73} $\alpha$-parameter, with observed accretion rates requiring $\alpha \sim 10^{-3}$--\,$10^{-2}$ \citep[e.g.][]{king07,rafikov17}. The final dispersal of the disc, at late times, is inconsistent with viscous accretion, and is usually attributed to photoevaporative winds \citep{hollenbach94,alexander06a,owen10}. This picture of (gas) disc evolution has been explored through a large body of both observational and theoretical work (see \citealt{rda_pp6} and \citealt{ep17}, and references therein).

However, it has also long been recognised that large regions of protoplanetary discs are insufficiently ionized to couple well to magnetic fields \citep{gammie96}. In this regime non-ideal magnetohydrodynamic (MHD) effects dominate, and act to suppress the MRI \citep[e.g.,][]{armitage11}. The resulting non-zero magnetic flux invariably drives a magnetised disc wind \citep[e.g.,][]{si09,fromang13,bai13b,bai13,gressel15}, whose properties are primarily determined by the magnetic field rather than the local disc conditions \citep[e.g.,][see also the review by \citealt{lesur_pp7}]{lesur21}. Magnetised winds carry both mass and angular momentum away from the disc, leading to a scenario where disc accretion is instead driven by the wind \citep[e.g.,][]{salmeron11}. We therefore have two competing pictures of protoplanetary disc accretion (viscous or wind-driven), which can both -- at least in broad terms -- successfully reproduce the demographics of observed disc populations \citep[e.g.,][]{lodato17,somigliana20,tabone22a,tabone22b}.

Observations do not currently give a clear picture of whether turbulent or wind-driven accretion is dominant. Close to the star ($\lesssim $0.1\,AU), where thermal ionization is sufficient to drive the MRI, observations of both turbulent velocities \citep{carr04} and the bulk properties of the disc \citep{mcclure19} imply $\alpha \gtrsim 10^{-2}$, easily large enough to account for the observed stellar accretion rates. At larger radii ($>$10\,AU), by contrast, observations of turbulent velocities and dust settling both typically yield much lower values, $\alpha \lesssim 10^{-3}$ \citep{flaherty18,flaherty20,teague18,dullemond18}, and the apparent lack of viscous spreading implies a similarly low $\alpha$ \citep{trapman20,long22}. 
Magnetised winds with high mass-loss rates are detected through both molecular \citep[e.g.,][]{devalon20,booth21} and atomic \citep[e.g.,][]{banzatti19,whelan21} tracers, while in other systems we see clear evidence of photoevaporative mass-loss \citep[e.g.,][]{pascucci11}, especially from more evolved discs \citep[][]{pascucci20}. However, how the mass-loss in these winds varies with both radius and time remains highly uncertain \citep{pascucci_pp7}. The dominant driver of disc accretion therefore remains unknown. 

Demographic studies have traditionally been our primary tool for understanding disc evolution on $\sim$Myr time-scales \citep[e.g.,][]{haisch01,aw05,fedele10}. However, the global disc properties used in these studies -- disc masses and stellar ages in particular -- are still plagued by large systematic uncertainties \citep{soderblom_pp6,miotello_pp7}. The disc accretion rate on to the star can be measured directly from observed accretion luminosities \citep[e.g.,][]{hartmann16}. Accretion measurements are still subject to significant uncertainties, most notably the bolometric corrections \citep[e.g.,][]{pittman22}, and the effects of short time-scale variability \citep[e.g.,][]{venuti17}. However, with {\it Gaia} now providing accurate stellar distances, accretion rates have become the best-determined of these demographic indicators \citep{manara_pp7}. Here we propose that the distribution of observed accretion rates can be used as a stand-alone diagnostic of protoplanetary disc evolution, and show that it can distinguish clearly between viscous and wind-driven accretion.


\section{A tale of two disc models}\label{sec:models}

Our statistical approach is relatively simple: in order to avoid the myriad of systematic uncertainties associated with stellar ages, disc masses, and other inferred observables \citep[e.g.][]{andrews20}, we limit our analysis to considering only the {\it distribution} of accretion rates. To do this we make a single simplifying assumption: that the observed accretion rates are representative of the underlying distribution. Essentially we assume that the dispersion in the evolutionary states of the discs is large enough that the full accretion histories are well-sampled.  With this assumption in place we need only consider the observed distribution of accretion rates, as any model for $\Mdot(t)$ can be inverted to give a probability distribution function $p(\Mdot)$.  By ``marginalising'' over time in this manner, we are able to perform a more detailed statistical analysis than has previously been possible. 

In order to test this approach we consider two models for protoplanetary disc evolution: a viscous model, where accretion is driven by disc turbulence; and a wind-driven model, where the disc accretes due to torques from a magnetised wind. We describe each of these models in turn below.


\subsection{Viscous / photoevaporation model}\label{sec:viscous}
Our viscous model assumes that the disc evolves subject to turbulent transport of angular momentum (``viscosity''), and mass-loss due to photoevaporation \citep[e.g.,][]{clarke01,alexander06b,owen10,picogna19}. We use the similarity solution of \citet[][see also \citealt{hartmann98}]{lbp74}, and adopt the Green's function solution of \citet{ruden04} for the effects of photoevaporation. This is a somewhat simplified approach, but in practice gives a functional form of $\Mdot(t)$ that is consistent with more sophisticated models. 

The similarity solution assumes a time-independent, power-law form for the disc viscosity $\nu$ as a function of radius $R$, 
\begin{equation}
\nu \propto R^{\gamma} \, ,
\end{equation}
and results in an accretion rate that evolves as
\begin{equation}\label{eq:ss}
\Mdotss(t) = \frac{M_{\mathrm d,0}}{2(2-\gamma)t_{\nu}} \tau^{-\frac{5/2 - \gamma}{2-\gamma}} \, .
\end{equation}
Here $M_{\mathrm d,0}$ is the initial disc mass, and $t_{\nu}$ is the viscous scaling time of the similarity solution. The first term therefore represents the initial accretion rate
\begin{equation}
\Mdot_0 = \frac{M_{\mathrm d,0}}{2(2-\gamma)t_{\nu}} \, ,
\end{equation}
and the dimensionless time $\tau$ is given by
\begin{equation}
\tau = \frac{t}{t_{\nu}}+1 \, .
\end{equation}
$\Mdotss(t)$ therefore follows a power-law form for $t \gg t_{\nu}$, and as long as $\Mdot_0$ significantly exceeds the highest observed value the probability distribution $p(\Mdotss)$ depends only on the power-law index $\gamma$. Observed disc surface density profiles, and demographic studies, both suggest that plausible values of $\gamma$ range from $\simeq 0.5$--1.5 \citep[e.g.][]{andrews09,andrews10,zhang17,lodato17}.

To capture the late-time behaviour (when photoevaporation leads to the cessation of accretion) we modify the similarity solution by introducing a polynomial ``cut-off'' (following \citealt{ruden04} \& \citealt{armitage07}), so that $\Mdot(t) \rightarrow 0$ as $t \rightarrow t_{\mathrm {max}}$:
\begin{equation}\label{eq:mdot_visc}
\Mdot(t) = \Mdotss(t) \left[1 - \left(\frac{t}{t_{\mathrm {max}}}\right)^{3/2}\right] \quad , \quad t \le t_{\mathrm {max}} \, . 
\end{equation}
With this prescription $t_{\mathrm {max}}$ is the disc lifetime, but in practice we do not use $t_{\mathrm {max}}$ as the second free parameter of the model. We instead define a ``cut-off'' accretion rate $\Mdot_{\mathrm c} = \Mdotss(t_{\mathrm {max}})$ as the free parameter: physically, $\Mdot_{\mathrm c}$ corresponds to the mass-loss rate due to photoevaporation. $\Mdot(t)$ therefore follows the similarity solution at early times, then drops rapidly to zero once the accretion rate falls below $\Mdot_{\mathrm c}$. This reproduces the behaviour of more sophisticated viscous/photoevaporation models \citep[e.g.,][]{aa09,picogna19} well, and with this form it is straightforward to invert $\Mdot(t)$ to find the probability distribution $p(\Mdot)$. 

Formally this model has four free parameters: $M_{\mathrm d,0}$, $t_{\nu}$, $\gamma$ and $\Mdot_{\mathrm c}$\footnote{Note that $p(\Mdot)$ does not depend strongly on the value of the polynomial index in Equation \ref{eq:mdot_visc}, so we do not vary it from the chosen value of 3/2.
}. However, the first two of these effectively just define the initial accretion rate $\Mdot_0 [ = M_{\mathrm d,0}/(2(2-\gamma)t_{\nu})]$, and the power-law nature of the model means that $p(\Mdot)$ is independent of $\Mdot_0$ as long as $\Mdot_0 \gg \Mdot_{\mathrm c}$. In practice $p(\Mdot)$ is therefore only sensitive to two parameters: the viscous power-law index $\gamma$ and the cut-off accretion rate $\Mdot_{\mathrm c}$.


\subsection{Wind-driven accretion model}\label{sec:wind}
For the case of wind-driven accretion we follow the recent models of \citet{tabone22a,tabone22b}. This framework allows for hybrid models in which viscous and wind-driven accretion both play a role, but here we consider the limiting case of purely wind-driven accretion. We adopt the solution from \citet{tabone22b}, where the accretion rate evolves as
\begin{equation}\label{eq:mdot_wind}
\Mdot(t) = \frac{M_{\mathrm d,0}}{2 t_{\mathrm {acc},0}(1+f_{\mathrm M})} \left(1 - \frac{\omega}{2 t_{\mathrm {acc},0}} t \right)^{-1 + \frac{1}{\omega}}
\end{equation}
Here $t_{\mathrm {acc},0}$ is the initial accretion time-scale (analogous to the viscous time-scale above), and $f_{\mathrm M}$ is the mass ejection-to-accretion ratio in the magnetised wind. $\omega$ is a dimensionless parameter between 0 and 1 which parametrizes the (unknown) dissipation of the disc's magnetic field with time ($\omega = 1$ corresponds to a constant magnetic field strength). The first term in Equation \ref{eq:mdot_wind} defines the initial accretion rate $\Mdot_0$, as in the viscous model, while the form of the second term sets the disc lifetime (i.e., the time at which $\Mdot(t)$ drops to zero) to be $2 t_{\mathrm {acc},0}/\omega$. 

This model therefore also has four free parameters: $M_{\mathrm d,0}$, $t_{\mathrm {acc},0}$, $f_{\mathrm M}$ and $\omega$. $f_{\mathrm M}$ serves only to change the conversion between disc mass and accretion rate in the initial conditions, which is at most an order-of-unity effect. However, the polynomial form of this model behaves differently to the viscous power-law, and here the probability distribution $p(\Mdot)$ is always sensitive to the initial accretion rate. The wind-driven model is therefore also primarily determined by just two parameters: the initial accretion rate $\Mdot_0 [=M_{\mathrm d,0}/(2t_{\mathrm {acc},0}(1+f_{\mathrm M}))]$, and the dissipation parameter $\omega$. 


\subsection{Comparison of the analytic models}\label{sec:comparison}


\begin{figure}
	\includegraphics[width=\columnwidth]{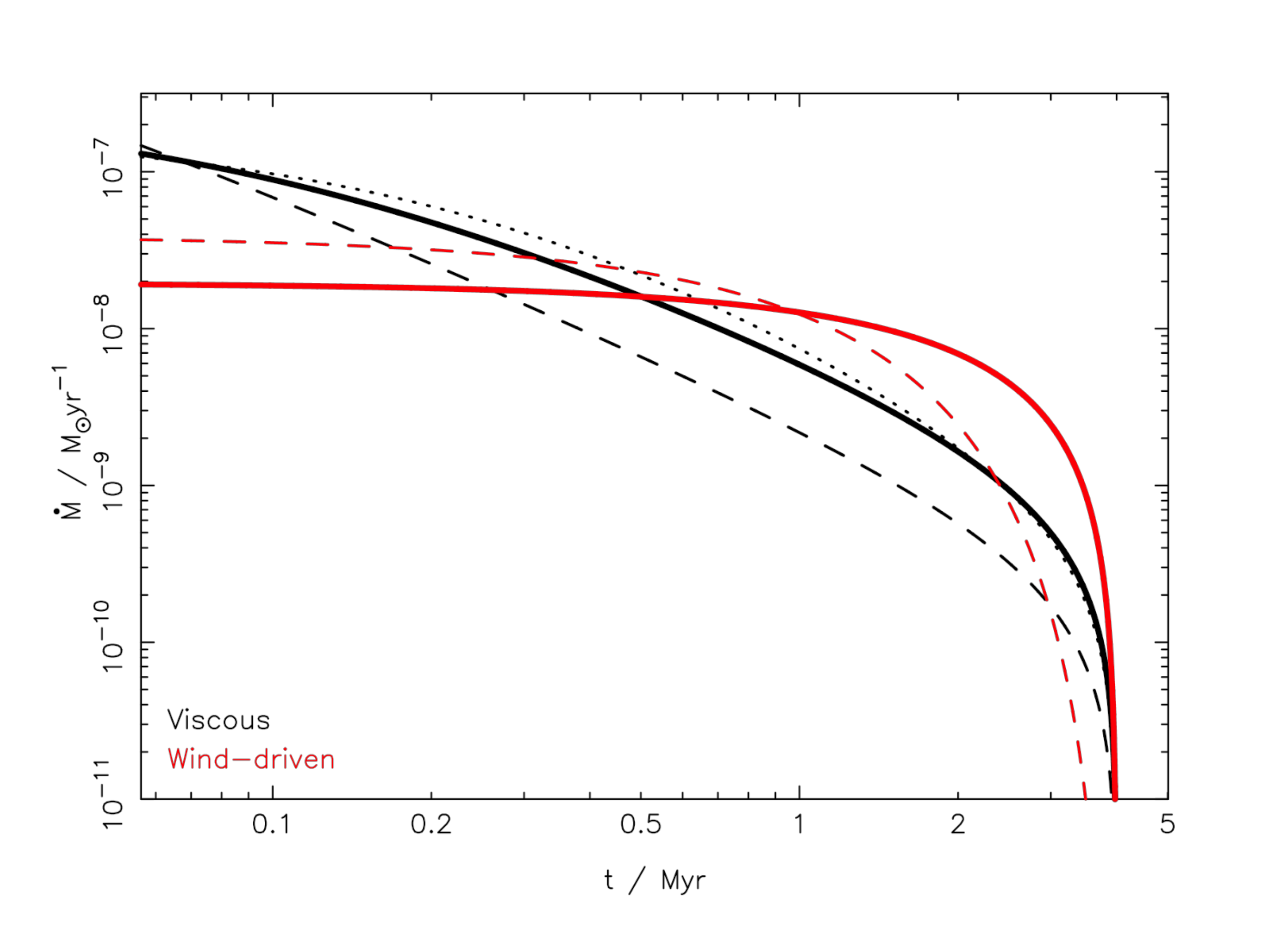}
        \vspace*{-12pt}
	\caption{Accretion rate as a function of time in the canonical viscous (black line; Equation \ref{eq:mdot_visc}) and wind-driven (red line; Equation \ref{eq:mdot_wind}) disc models. The viscous model is essentially a power-law, truncated below the cut-off rate $M_{\mathrm c}$, so most of the disc lifetime is spent at low accretion rates. By contrast, in the wind-driven model accretion declines geometrically, so most of the disc lifetime is spent at high accretion rates, close to the initial value $\Mdot_0$. The dashed and dotted lines show the effect of varying different model parameters (while keeping $M_{\mathrm d,0}$ and the disc lifetime fixed). The dashed black line shows a viscous model with a lower cut-off accretion rate ($\Mdot_{\mathrm c} = 3\times10^{-10}$\,\Msunyr), while the dotted black line shows a model with power-law index $\gamma = 3/2$. The dashed red line denotes a wind-driven model with $\omega = 0.2$ (which for a fixed disc lifetime also increases the initial accretion rate $\Mdot_0$).}
    \label{fig:mdot_t_analytic}
\end{figure}


Our first step is to compare the analytic forms of these models. We normalise both models to have the same initial disc mass $M_{\mathrm d,0} = 0.05$\,\Msun, and matching disc lifetimes of $t_{\mathrm {max}} = 2t_{\mathrm {acc},0}/\omega = 4$\,Myr. For this initial comparison we adopt canonical parameters of $\gamma = 1$ and $\Mdot_{\mathrm c} = 1\times10^{-9}$\,\Msunyr in the viscous model\footnote{For this comparison we have used the disc lifetime, $t_{\mathrm {max}}$, as an input parameter instead of the viscous time-scale. In this model this sets $t_{\nu} = 9.53\times10^4$\,yr, and therefore $\Mdot_0 = M_{\mathrm d,0}/2t_{\nu}=2.64\times10^{-7}$\,\Msunyr.}, and $f_{\mathrm M} = 0.6$ and $\omega = 0.4$ for the wind-driven model.

\begin{figure}
	\includegraphics[width=\columnwidth]{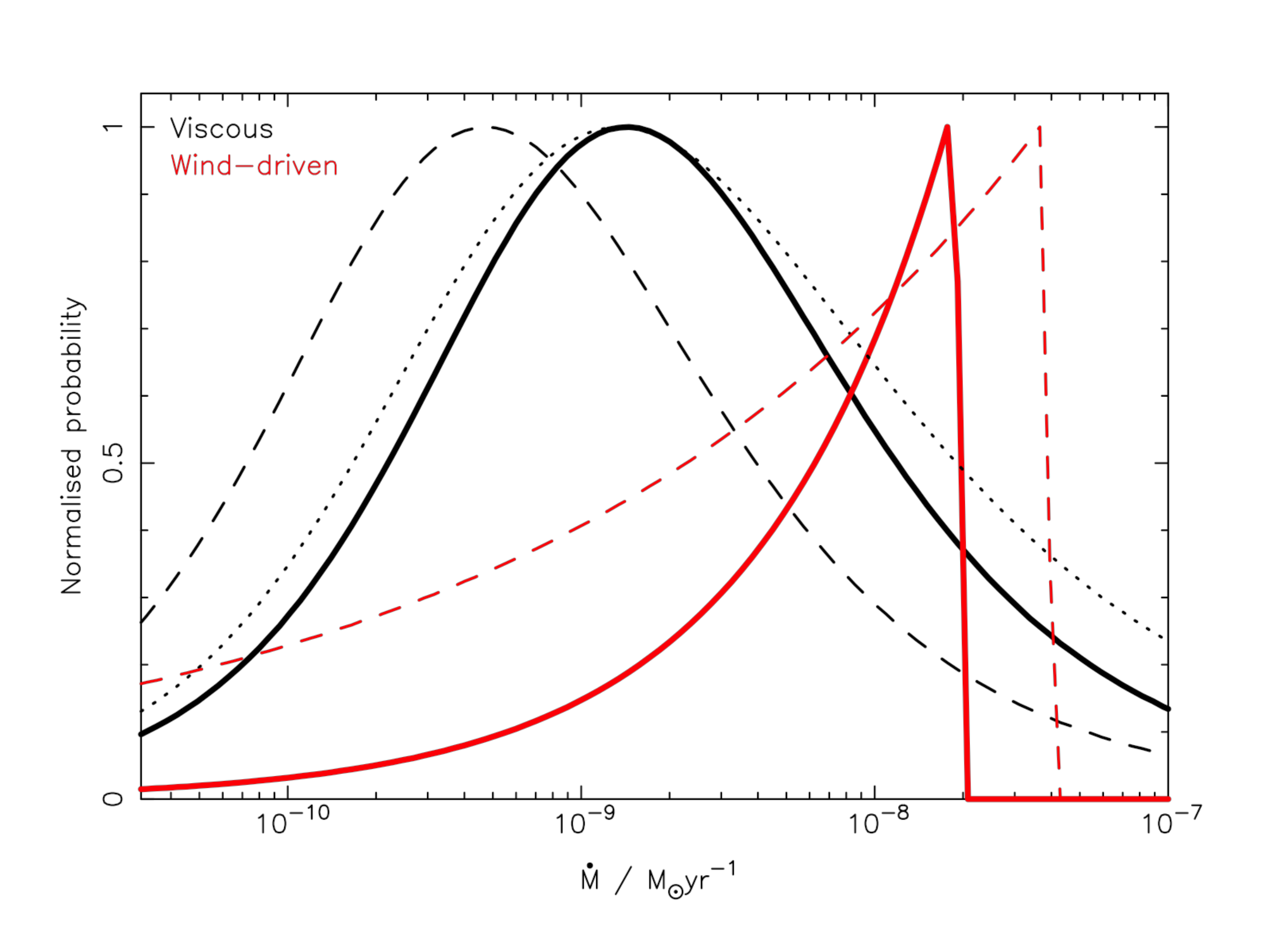}
        \vspace*{-12pt}
	\caption{Accretion rate probability distributions $p(\dot{M})$ for the two disc models, assuming uniform sampling of $\dot{M}(t)$. As in Fig.\ref{fig:mdot_t_analytic}, the solid black line represents the viscous model and the solid red line the wind-driven model. For the viscous model the distribution peaks close to the cut-off rate $M_{\mathrm c}$, and is approximately symmetric around this peak. By contrast, the distribution for the wind-driven model always peaks at the initial (maximum) accretion rate $\dot{M}_0$, and declines to smaller values. As in Fig.\,\ref{fig:mdot_t_analytic}, the dashed and dotted curves show the effect of changing the model parameters. In the viscous model, lowering the cut-off rate $M_{\mathrm c}$ shifts the peak of the distribution to lower $\Mdot$ by the same factor, while changing the power-law index $\gamma$ has only a minor impact. In the wind-driven model, a lower value of the dissipation parameter $\omega$ results in a broader distribution, and also shifts the peak to higher $\Mdot$ (as for a fixed disc lifetime, $\Mdot_0 \propto t_{\mathrm {acc},0}^{-1} \propto \omega^{-1}$).}
    \label{fig:p_mdot_analytic}
\end{figure}


Fig.\,\ref{fig:mdot_t_analytic} shows how $\Mdot$ varies as a function of time in these two canonical models, as well as the effects of varying the key model parameters. In the viscous model $\Mdot(t)$ is a power-law truncated at low accretion rates, so the majority of disc lifetime is spent at low $\Mdot$, close to the cut-off value $\Mdot_{\mathrm c}$. By contrast, the geometric decline in the wind-driven model sees the disc spend most of its lifetime at high $\Mdot$, close to the initial accretion rate $\Mdot_0$. This behaviour is reflected in the resulting distribution functions $p(\Mdot)$, shown in Fig.\,\ref{fig:p_mdot_analytic}. The distribution of accretion rates for the viscous model is peaked close to $\Mdot_{\mathrm c}$, with power-law declines to higher and lower values; while the distribution for the wind-driven model peaks at the limiting initial accretion rate $\Mdot_0$, and declines as a power-law to lower values. The form of $p(\Mdot)$ in the viscous model is insensitive to $t_\nu$ as long as $t_{\mathrm {max}} \gg t_{\nu}$ (or, equivalently, $\Mdot_0 \gg \Mdot_{\mathrm c}$), and is only weakly sensitive to the viscous power-law index $\gamma$ (which changes the slope of the decline to high $\Mdot$). For any plausible choice of $\gamma$ we find a broad distribution of accretion rates that is close to symmetric [in $\log(\Mdot)$] around the peak. In the wind-driven model we see that $p(\Mdot)$ has an upper cut-off set by the initial accretion rate (which depends primarily on $t_{\mathrm {acc},0}$), while the width of the distribution (or alternatively the slope of the decline to low $\Mdot$) is determined by $\omega$ (indeed, for $\omega=1$, $p(\Mdot)$ is a $\delta$-function). Low values of $\omega$ ($\lesssim 0.2$) can give a comparably broad distribution to the viscous model, but $p(\Mdot)$ in the wind-driven model is always asymmetric, and peaks at the maximum value $\Mdot_0$\footnote{This also implies that significant scatter in $\Mdot_0$ is required for wind-driven accretion to reproduce the full range of observed accretion rates, which extend up to $\sim10^{-6}$\,\Msunyr (see Fig.\,\ref{fig:pp7_histogram}.)}. These two models therefore predict qualitatively and quantitatively distinct distributions of accretion rates. In the next sections we consider how large an observed sample is required to distinguish between these models, and how this is affected by scatter in the various model parameters. 

\subsection{Generating simulated data}\label{sec:fake_data}


\begin{figure}
       \includegraphics[width=\columnwidth]{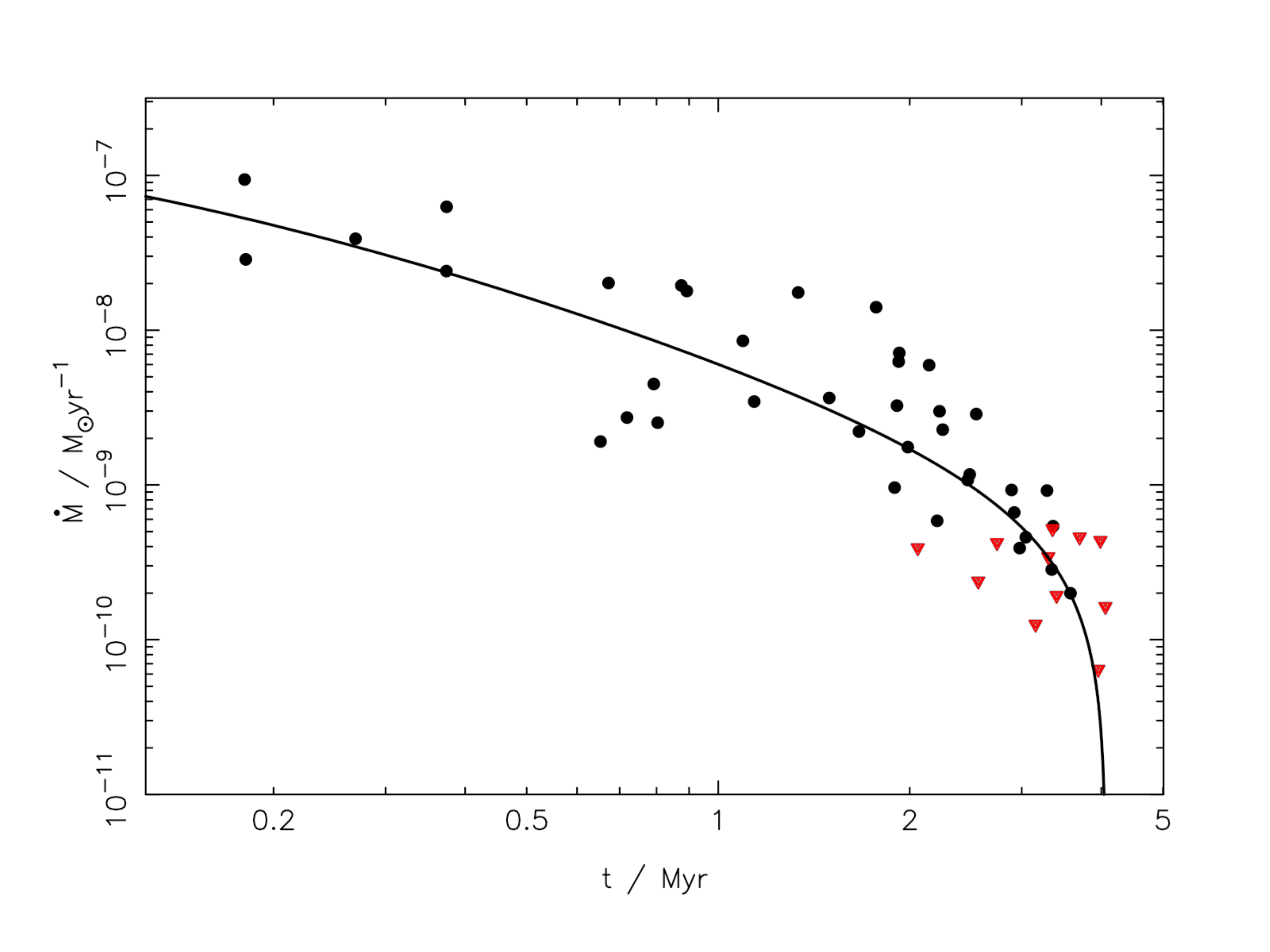}       
       \vspace*{-12pt}
       \caption{Simulated data generated for a model with $\gamma=1$, $\Mdot_{\mathrm c} = 10^{-9}$\,\Msunyr, and $\Delta_{\mathrm M}=0.35$.  The solid curve shows the analytic form of $\Mdot(t)$, and the points show the $N=50$ simulated ``observations'' generated from this curve, using the procedure described in Section \ref{sec:fake_data}.  Black circles represent detections; red triangles represent upper limits.}     
           \label{fig:fake_data}
\end{figure}


To understand how we can distinguish between these models observationally, we first generate simulated distributions of accretion rates. Our method of generating simulated observations is as follows.  We first define a model $\Mdot(t)$, valid over a range $[0,t_{\mathrm {max}}]$.  We then randomly sample $N$ values of $t_i \in [0,t_{\mathrm {max}}]$, and compute a set of $N$ values $\Mdot_i(t_i)$. To simulate a realistic set of observations we then modify the sample $\Mdot_i$ in two ways, first adding a random scatter to the data, and then using a selection function to designate a sub-set of the data points as upper limits (i.e., non-detections).  The scatter accounts for both real effects (such as variability), and also for observational uncertainties.  The scatter is applied in log-space: the modified values $\Mdot_i' $ are computed as
\begin{equation}
\log_{10}(\Mdot_i') = \log_{10}(\Mdot_i) + \delta_{\mathrm M,i} \, ,
\end{equation}
where the (log-)scatter in the individual points, $\delta_{\mathrm M,i}$, is drawn randomly from a zero-mean Gaussian distribution with standard deviation  $\Delta_{\mathrm M}$.  Following \citet{manara_pp7}, we adopt $\Delta_{\mathrm M} = 0.35$\,dex as the magnitude of this ``observational'' scatter. We then designate values as either detections or upper limits according to a simple exponential selection function:
\begin{equation}
p_{\mathrm {det}}(\Mdot) = \left\{ \begin{array}{ll}
1 & \textrm{if} \, \Mdot \ge \Mdot_{\mathrm t} \\
\exp\left(\left[\log_{10}(\Mdot)-\log_{10}(\Mdot_{\mathrm t})\right]/\sigma_{\mathrm M}\right) & \textrm{if} \, \Mdot < \Mdot_{\mathrm t} \\
\end{array} \right.
\end{equation}
Here $p_{\mathrm {det}}$ is the probability of detection.  Based on the relative numbers of detections and upper limits in real data (\citealt{ingleby11,manara_pp7}; see also Fig.\,\ref{fig:pp7_histogram}), we set the threshold accretion rate (above which all data points are detections) to be $\log_{10}(\Mdot_{\mathrm t}) = -9.25$, and $\sigma_{\mathrm M} = 0.5$.  Individual data points are then designated as either detections or upper limits by sampling randomly from the distribution $p_{\mathrm {det}}(\Mdot_i')$.  If a data point is designated as an upper limit, it is then assigned a final value $\Mdot_j$ by sampling randomly (in log-space) from the range $[\Mdot_i',\Mdot_{\mathrm t}]$ (i.e., we assign an ``observed'' upper limit which lies between the true value and the detection threshold).  An example of this procedure is shown in Fig.\,\ref{fig:fake_data}. Our procedure is somewhat simplified, and neglects the fact that in real observations the detection threshold for the accretion luminosity depends on both the stellar mass and age \citep[e.g.,][]{manara17}. However, for a given stellar mass the dependence on age (which is due to the decreasing stellar luminosity) is not very strong, so these simplifications are reasonable. The outcome of this process is a set of $N$ accretion rates, with both scatter and detection limits that are broadly representative of real observations.


\subsection{Distinguishing between the models}\label{sec:ks_tests}

\begin{figure}
       \includegraphics[width=\columnwidth]{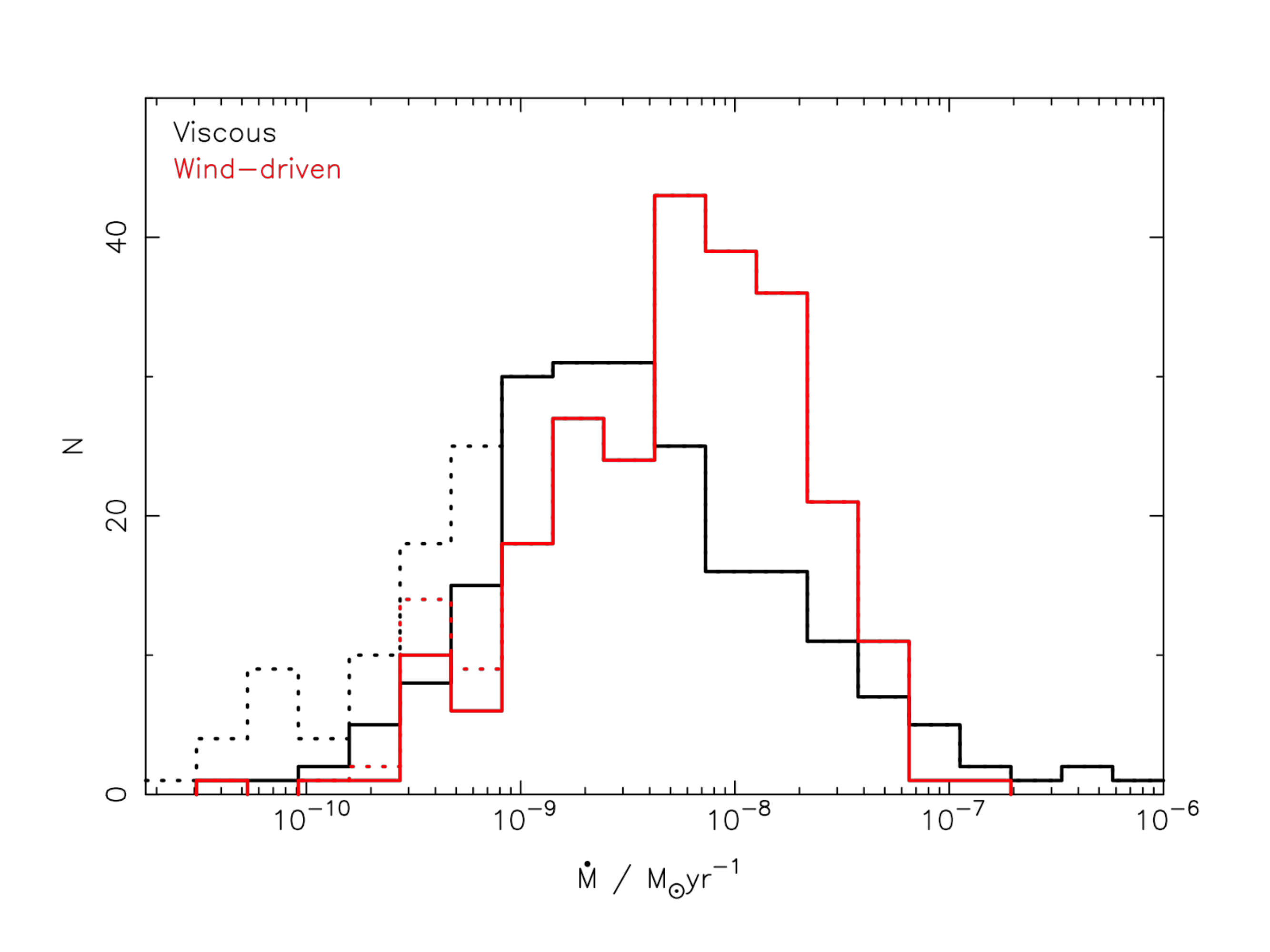}      
       \vspace*{-12pt}
       \caption{Histograms of simulated data generated from the canonical disc models, using $N=250$ sources. The viscous (black histogram) and wind-driven (red histogram) models use the same parameters as in Figs.\,\ref{fig:mdot_t_analytic} \& \ref{fig:p_mdot_analytic}, with ``observational'' scatter of $\Delta_{\mathrm M}=0.35$\,dex. The solid histograms show only the simulated detections, while the dotted histograms also include the upper limits. Despite the scatter, the distributions of accretion rates from the two models are clearly distinguishable. The probability that the two samples (of detections) are drawn from the same underlying distribution is $1.1\times10^{-5}$.}
             \label{fig:fake_histogram}
\end{figure}
     

The simplest question we can now ask is whether or not a sample of accretion rates can distinguish between these two disc evolution models and, if so, how large a sample is required. We initially draw samples of $N$ simulated accretion rates from each model, following the procedure described in Section \ref{sec:fake_data}; an example is shown in Fig.\ref{fig:fake_histogram}. For this initial comparison the parameters of both models are fixed to the canonical values given in Section \ref{sec:comparison}; the only differences between this comparison and that in Fig.\ref{fig:p_mdot_analytic} is the finite sampling, and the introduction of the ``observational'' scatter. The distribution accretion rates from the viscous model has a mean (in $\log_{10}(\Mdot)$) of $-8.7$ and a standard deviation of 0.83\,dex; the distribution from the wind-driven model has a mean of $-8.3$ and a standard deviation of 0.67\,dex. The wind-driven model also produces a notably asymmetric distribution, with a long ``tail'' to low $\Mdot$. The distributions still peak at the same values as in the analytic models ($\Mdot_{\mathrm c}$ for the viscous model, and $\Mdot_0$ for the wind-driven model), and in both cases the intrinsic width of the distribution significantly exceeds the observational scatter (0.35\,dex). Given the highly inhomogeneous nature of the upper limits on $\Mdot$ in real observations, we consider only the detections when comparing our samples quantitatively\footnote{With these simulated data, including the upper limits in the analysis substantially increases its statistical power. However, in real data the upper limits are usually very inhomogeneous (see Section \ref{sec:sample}), and are therefore of limited use in practice.}. We then perform a Kolmogorov-Smirnov (KS) test to find the probability that the two sets of $\Mdot$ values were drawn from the same underlying distribution. For the example shown in Fig.\ref{fig:fake_histogram}, using $N=250$, the KS probability is $1.1\times10^{-5}$, so we are able to distinguish between the models at high confidence.

We generalise this procedure by repeating this process for a range of values of $N$. There is significant stochasticity in the results, especially at small $N$, so for each value $N$ we repeat this process 1000 times. The resulting distribution of K-S probabilities is plotted in Fig.\ref{fig:all_the_ks_probs}: the median value (as a function of $N$) is denoted by the black line, while the shaded regions denote the 25th and 75th percentiles of the distribution. We see for small sample sizes the distributions of $\Mdot$ from the two models are usually similar, but for $N\gtrsim 100$ the accretion rate distribution allows us to distinguish between them at high confidence.


\subsection{A more realistic comparison}\label{sec:ks_tests_scatter}
Despite the inclusion of ``observational'' scatter, however, this remains a highly idealised comparison, as the two models have quite different functional forms and a fixed set of parameters. A more realistic comparison is to consider models where the input parameters span broad ranges, as suggested by demographic studies \citep{somigliana20,tabone22b}. We adopt a pessimistic set of assumptions here, maximising the plausible spread in the model parameters to make a stringent test. We therefore apply scatter to our model parameters as follows\footnote{This procedure results in a slight inconsistency between how the two models are treated: in the wind-driven model the range of lifetimes is prescribed, while no limit on $t_{\nu}$ is imposed on the viscous model. This makes no difference here, as we consider only the distribution of $\Mdot$ (effectively integrating over the disc lifetimes), but we note that the extremes of our parameter space include some models with unrealistically long or short viscous time-scales.}:
\begin{itemize}
\item $M_{\mathrm d,0}$ -- the initial disc mass (in both models) is drawn from a log-normal distribution with a mean of $0.03$\,\Msun\ and a standard deviation of 0.5\,dex.
\item $\Mdot_{\mathrm c}$ -- the cut-off accretion rate in the viscous model is also drawn from a log-normal distribution, with a mean of $10^{-9}$\,\Msunyr\ and a standard deviation of 1.0\,dex.
\item $\gamma$ -- the viscous power-law index is drawn from a uniform distribution spanning the range $[0.5,1.5]$.
\item $\omega$ -- the magnetic dissipation parameter is drawn from a Gaussian distribution with a mean of 0.4 and a standard deviation of 0.2. We additionally impose a minimum value of $\omega = 0.1$, as very small values of $\omega$ lead to unphysical results. 
\item $t_{\mathrm {acc},0}$ -- following \citet{tabone22b}, the accretion time-scale is drawn from an exponential distribution $\exp(-t/\tau)$, with $\tau = 2.5$\,Myr. This sets the characteristic initial accretion rate $M_{\mathrm d,0}/\tau = 1.2\times10^{-8}$\,\Msunyr.
\end{itemize}

To generate simulated data with this scatter we draw a single accretion rate for each set of model parameters, and repeat this process $N$ times to generate the sample of ``observed'' accretion rates for each model. This is analogous to observing $N$ different model discs, with a random set of parameters, at random times in their evolution, and this ``intrinsic'' scatter in the model parameters dominates over the ``observational'' scatter described in Section \ref{sec:fake_data} (though the latter is still applied).

\begin{figure}
\centering
       \includegraphics[width=\columnwidth]{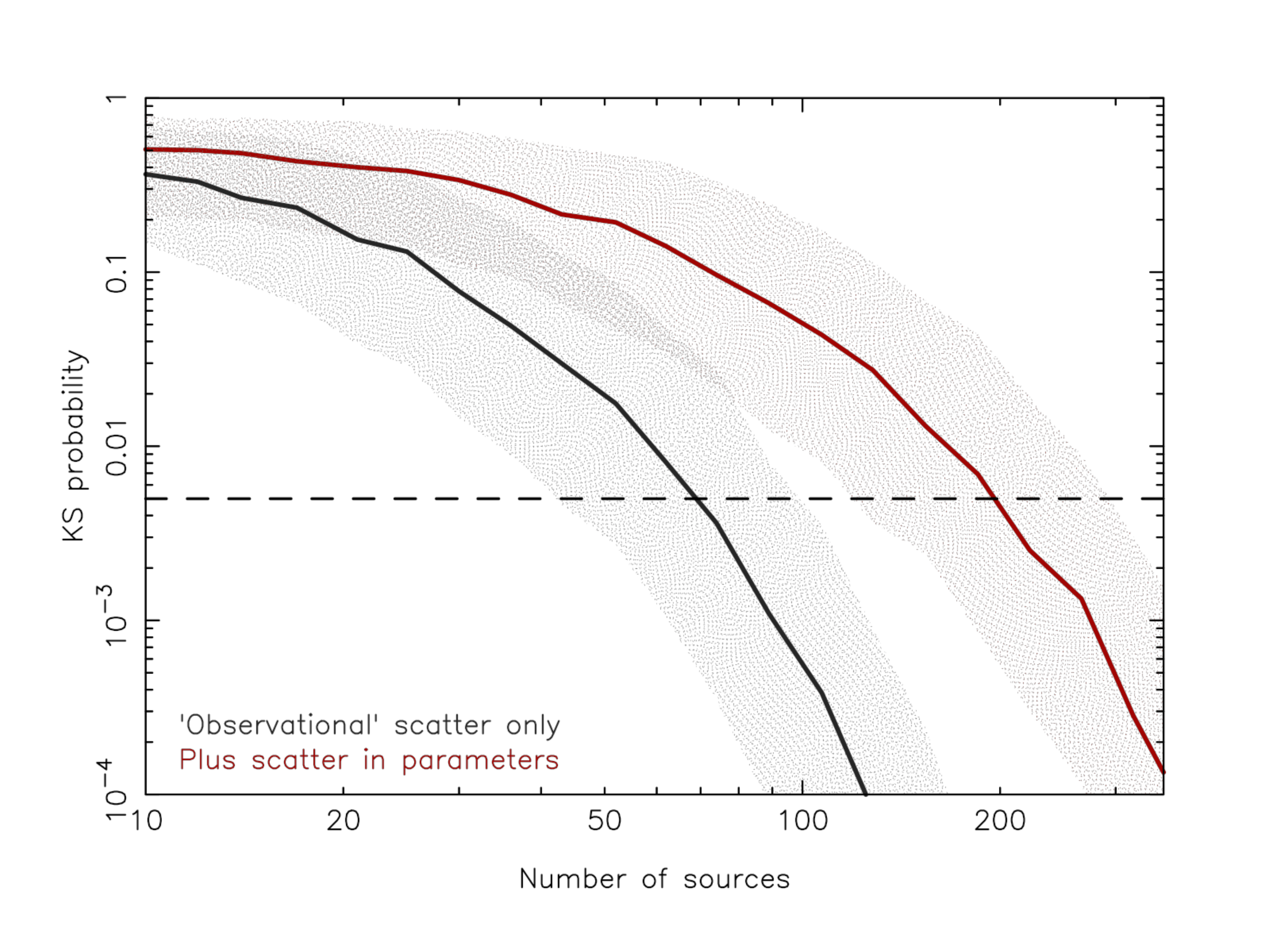}      
       \vspace*{-12pt}
       \caption{Probability that samples of accretion rates from the two different disc models could have been drawn from the same underlying distribution, as a function of the sample size $N$. Black/grey represents the ``basic'' models, with only observational scatter in $\dot{M}$, while red denotes the results with significant scatter applied to all the input parameters. For each value of $N$ the KS test was repeated for 1000 different random realisations of the models; the lines represent the median KS probability, while the shaded areas span the 25th to 75th percentiles of the distribution. The dashed horizontal line marks a probability of 0.5\%. For the basic model $N\gtrsim100$ is sufficient to distinguish between viscous and wind-driven accretion, but with scatter applied to the input parameters the required sample size rises to $N\gtrsim300$.}     
           \label{fig:all_the_ks_probs}
\end{figure}

We see from Fig.\ref{fig:all_the_ks_probs} that introducing scatter in the parameters makes it significantly more difficult to discriminate between the two models. Nevertheless, $N \gtrsim 300$ is still sufficient to distinguish between the two models at high confidence. The fact that a relatively modest sample size can still separate these models clearly even when we make very pessimistic assumptions about the model parameters (i.e., 1--2 orders-of-magnitude scatter) is very encouraging. We therefore conclude that the distribution of accretion rates can provide significant insight into protoplanetary disc physics, and has the potential to discriminate cleanly between viscous and wind-driven disc accretion. 


\section{Comparisons with observed accretion rates}

\subsection{The sample}\label{sec:sample}
We now seek to test our method using real accretion rate observations, and for this we use the compilation of data from \citet{manara_pp7}. Their complete sample contains 865 discs, of which 288 have measured accretion rates. However, the sample covers a wide range in stellar mass, so the global distribution of accretion rates is dominated by the well-known $\dot{M} \propto M_*^2$ trend \citep[e.g.,][]{muzerolle05,mohanty05,manara17}. To use the accretion rate distribution as a test of disc evolution we must therefore consider a restricted range in stellar mass. Formally the solutions in Section \ref{sec:models} do not depend on the stellar mass, but our model parameters are based on studies of T Tauri stars with masses $\simeq 0.5$--1\,\Msun. We therefore limit our analysis stellar masses in the range 0.3--1.2\,\Msun\ (i.e., 0.6\,\Msun, plus or minus a factor of 2). This leaves a sample of 121 objects with accretion rate measurements, of which 100 are detections and 21 are upper limits. Modest variations in this range in stellar mass do not alter our results significantly, but extending the range to $\lesssim 0.2$\,\Msun\ sees the distribution dominated by the stellar mass trend. 

\begin{figure}
       \includegraphics[width=\columnwidth]{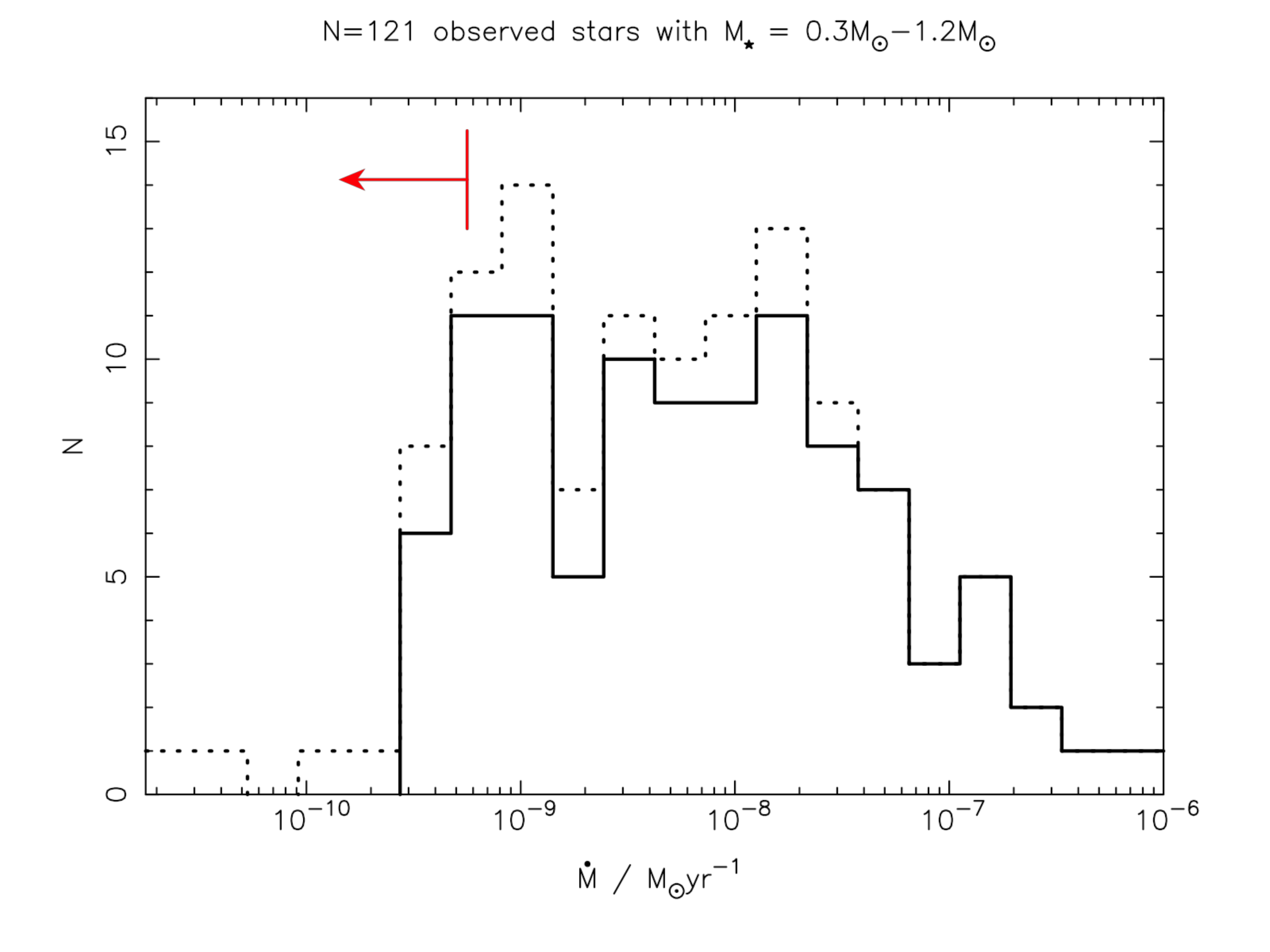}      
       \vspace*{-12pt}
       \caption{Distribution of observed accretion rates from the compilation of data in \citet{manara_pp7}: the solid line shows only detections, while the dashed histogram shows both detections and upper limits. We restrict our analysis to the traditional ``T Tauri star'' range in stellar mass, defined here as 0.3--1.2\Msun, which results in a sample of 121 sources. However, modest changes in the range of stellar masses we consider do not change the distribution significantly. The red arrow denotes the detection threshold, $\Mdot_{\mathrm t}$, applied to our simulated data (as described in Section \ref{sec:fake_data}.)}     
           \label{fig:pp7_histogram}
\end{figure}

The resulting distribution of accretion rates, for 121 discs, is shown in Fig.\,\ref{fig:pp7_histogram}. Several points about the distribution are notable. First, the gradual decline to high $\Mdot$ is broadly consistent with the canonical viscous model (see Fig.\,\ref{fig:p_mdot_analytic}), but inconsistent with the sharp cut-off predicted by the canonical wind-driven model. Reproducing the observed data with wind-driven accretion therefore requires significant scatter in the input parameters.

By contrast, at the low-$\Mdot$ end of the distribution the cut-off is fairly sharp, with no detections (and only 4 upper limits) below $3 \times 10^{-10}$\,\Msunyr. This primarily reflects observational detection limits, and physically corresponds to the level at which the accretion luminosity can no longer be readily detected above the (very bright) chromospheric emission from T Tauri stars \citep{ingleby11,manara13}. New accretion tracers (such as He\,{\sc i}) have pushed the detection threshold significantly lower \citep{atom22,atom23}, but these have not yet been applied to large samples. 

Finally, the upper limits in the \citet{manara_pp7} sample span more than two orders of magnitude, and are clearly not homogeneous. Given this, the relatively small number of upper limits in the sample, and the fact that these data are compiled from surveys which generally exclude weak-lined T Tauri stars and low accretors, the statistical significance of these upper limits is questionable. As a result we exclude the upper limits from our subsequent analysis, and from here onwards (for both models and data) consider only the detected accretion rates.


\subsection{Statistical analysis}\label{sec:stats}
We saw in Section \ref{sec:ks_tests_scatter} that a sample size of $N \gtrsim 300$ is required to distinguish strongly between the viscous and wind-driven accretion models. With only 100 detections we therefore do not expect the sample of accretion rates from \citet{manara_pp7} to be large enough for this purpose, and this is indeed what we find. The ``tail'' of the observed distribution at high $\Mdot$ means that the viscous model is weakly favoured, but in both cases the canonical models are broadly consistent with the data (a KS test fails to exclude either model). Moreover, given the inhomogeneous nature of the sample, and in particular the lack of a homogeneous set of upper limits, any statistical conclusions will inevitably be dominated by these uncertainties. As a result we do not pursue a more sophisticated statistical approach (such as MCMC) to constrain the model parameters. However, a simpler analysis still yields some interesting results. 

\begin{figure}
\centering
       \resizebox{\hsize}{!}{
       \includegraphics[angle=0]{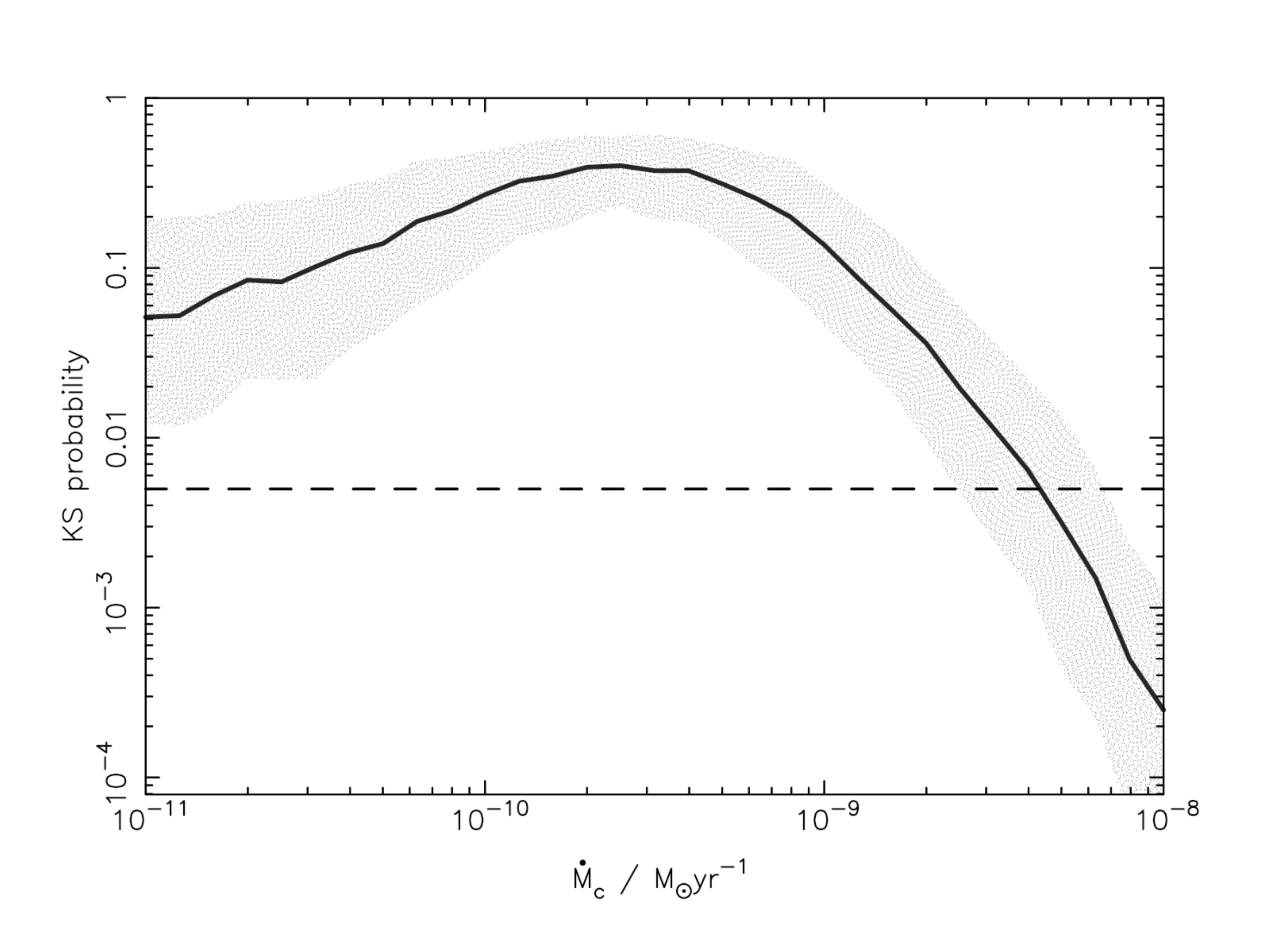}
       }
       \vspace*{-12pt}
       \caption{Probability that the observed accretion rates from \citet{manara_pp7} are drawn from the same underlying distribution as our viscous model (Equation \ref{eq:mdot_visc}), as a function of the cut-off accretion rate $\Mdot_{\mathrm c}$. As in Fig.\,\ref{fig:all_the_ks_probs}, the line represents the median from 1000 random realisations of the KS test, while the shared area spans the 25th to 75th percentiles of the distribution. The dashed horizontal line again marks a probability of 0.5\%. We see that values of $\Mdot_{\mathrm c}$\,$>$\,$5\times10^{-9}$\,\Msunyr\ are strongly ruled out by our analysis: if photoevaporation terminates disc accretion, then the photoevaporative mass-loss rates must be low.}     
       \label{fig:mdot_c}
\end{figure}

In order to place (weak) constraints on the model parameters using the framework described in Section \ref{sec:ks_tests_scatter}, we repeat the analysis (with $N=100$) while holding a single parameter fixed. We repeat the KS test for 1000 random realisations of the model for each value of the fixed parameter, and study how the probabilities vary\footnote{This is effectively a crude way of marginalising over the multi-parameter space to constrain a single parameter.}. In most cases we do not place any meaningful constraints on the model parameters, but we do recover two notable results. In the viscous model the power-law index $\gamma$ is not usefully constrained, but the large number of observed discs with accretion rates $\lesssim 10^{-9}$\,\Msunyr\ is inconsistent with higher values of the cut-off rate $\Mdot_{\mathrm c}$. We do not place any lower limit to $\Mdot_{\mathrm c}$, but this is primarily due to the observational detection limits: there are simply not enough data points with $\dot{M} \lesssim 10^{-10}$\,\Msunyr\ to provide useful constraints at low $\Mdot_{\mathrm c}$. The variation of the KS probability with $\Mdot_{\mathrm c}$ is shown in Fig.\,\ref{fig:mdot_c}: we see that values of $\Mdot_{\mathrm c} \gtrsim 10^{-9}$\,\Msunyr are disfavoured, and values $>5\times10^{-9}$\,\Msunyr\ are strongly excluded. This suggests that if disc photoevaporation is responsible for the cessation of disc accretion, then the mass-loss rates in the photoevaporative winds must be $\lesssim 10^{-9}$\,\Msunyr. Improved characterisation of the low end of the $\Mdot$ distribution will provide a better measurement of this cut-off, and determine the photoevaporation rate accurately. 


\begin{figure}
\centering
       \resizebox{\hsize}{!}{
       \includegraphics[angle=0]{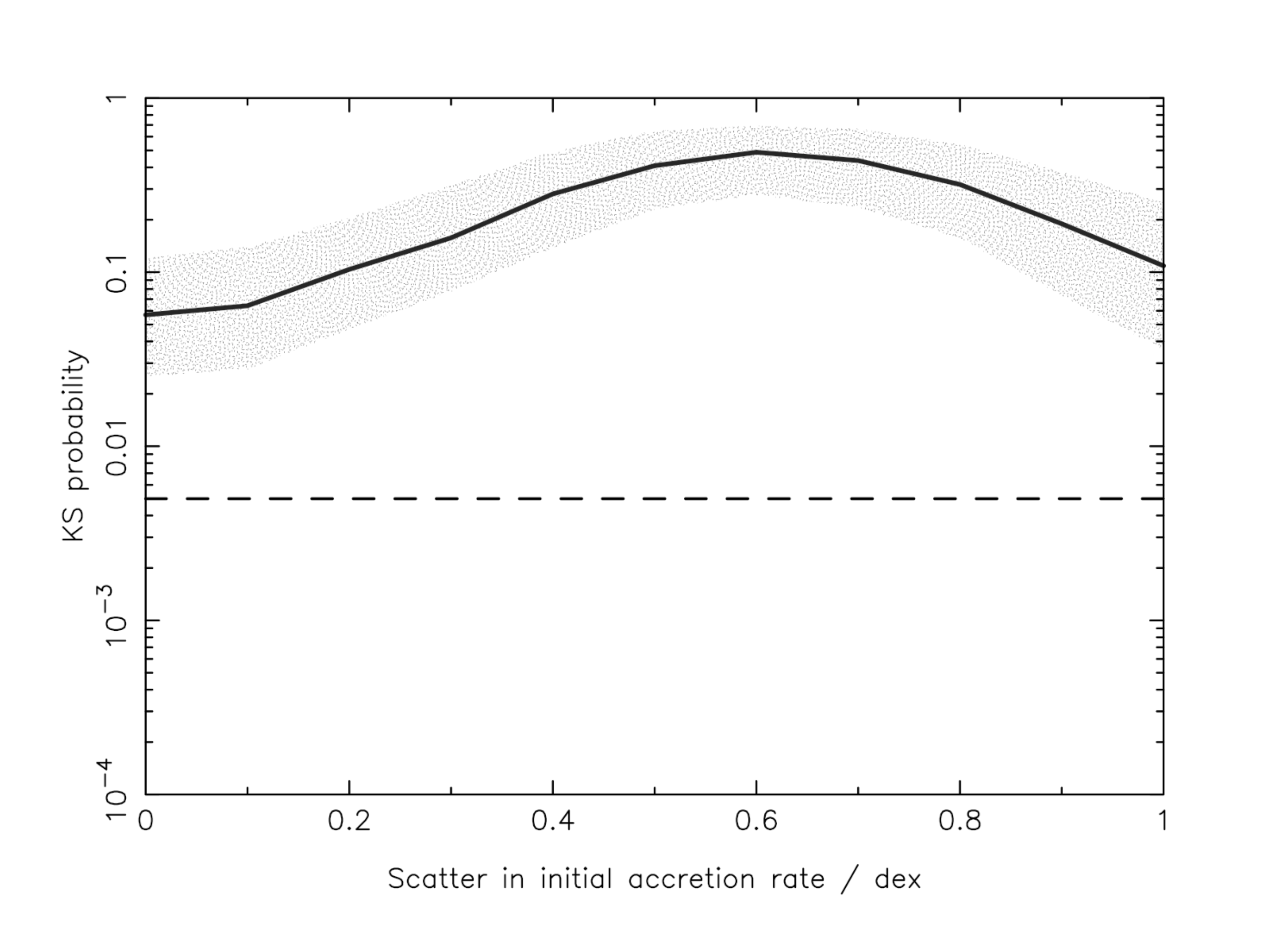}
       }
       \vspace*{-12pt}
       \caption{Probability that the observed accretion rates from \citet{manara_pp7} are drawn from the same underlying distribution as our wind-driven model (Equation \ref{eq:mdot_wind}), as a function of the scatter in the initial accretion rate. As in Figs.\,\ref{fig:all_the_ks_probs} \& \ref{fig:mdot_c}, the line represents the median from 1000 random realisations of the KS test, while the shared area spans the 25th to 75th percentiles of the distribution. The dashed horizontal line again marks a probability of 0.5\%. Although the probability peaks at a scatter of $\simeq 0.6$\,dex, none of the parameter space for the wind-driven models is ruled out.}     
       \label{fig:mdot0_scatter}
\end{figure}

For the wind-driven model the constraints are much weaker, and in fact none of our tests rule out any of the parameter space at $<$\,0.5\% probability (i.e., at the ``3-$\sigma$'' level). In order to reproduce the spread in accretion rates we require $\omega<1$ \citep[as found by][]{tabone22b}, and some scatter in the initial accretion rates is weakly favoured (see Fig.\,\ref{fig:mdot0_scatter}). There is also a weak preference for slightly lower initial accretion rates than in our canonical model, but given the inhomogeneity of the sample, and the degeneracies between the model parameters, we do not attach any statistical significance to this result. The wind-driven model is most strongly constrained by the upper end of the accretion rate distribution, and lowering the median value of $\Mdot_0$ requires increasingly large scatter to reproduce the highest observed accretion rates. However, this sensitivity to the initial conditions is hard to interpret, as at very early times the physical significance of these solutions is unclear. In real systems ``$t=0$'' is not well-defined, and the early evolution of protoplanetary discs is dominated by infall. Nevertheless, we are now able to measure samples of accretion rates during the Class I phase \citep[e.g.,][]{fiorellino23}, and the behaviour of the wind-driven models suggests that additional observations of high-$\Mdot$ discs may provide a useful test of wind-driven accretion. 

We therefore conclude that the compilation of accretion rates by \citet{manara_pp7} is broadly consistent with models of both viscous and wind-driven accretion. A modest expansion of the sample size (by a factor of 2--3) is needed in order to distinguish between these models clearly, and a homogeneous sample of upper limits would also increase the power of this method significantly. Expanding the sample of observed accretion rates for stars in the $\simeq 0.5$--1\,\Msun\ range to $\gtrsim300$ objects requires significant effort, and would represent most of the T Tauri stars in nearby ($\lesssim 150$\,pc) star-forming regions. Indeed, it seems unlikely that the sample of detections can easily by extended by this much. However, observations of the ``non-accreting'' weak-lined T Tauri stars are much more limited, and this may represent the most fruitful way to increase the sample size in the near future. New observations of low accretors are already yielding interesting results even from relatively small samples \citep[e.g.,][]{atom23}; a large, homogenous sample of accretion rates across both Class II and Class III discs would be a powerful statistical tool for understanding disc accretion.


\section{Discussion}

\subsection{Caveats and limitations}
We have shown that accretion rate statistics can distinguish clearly between two parametrized models of disc evolution, but the obvious question is whether or not the functional forms for $\Mdot(t)$ in Section \ref{sec:viscous} \& \ref{sec:wind} capture the underlying physical behaviour accurately. Mathematically, the difference in $p(\Mdot)$ seen in Fig.\,\ref{fig:p_mdot_analytic} can be understood by inspection of Equations \ref{eq:mdot_visc} and \ref{eq:mdot_wind}: the power-law form of the viscous model results in the disc spending most of its lifetime at low $\Mdot$; while the geometric form of the wind-driven model instead spends most of its lifetime at high $\Mdot$. The power-law form for the viscous model arises because the viscosity remains constant as the disc evolves, so $\Mdot \propto \Sigma$. The accretion rate therefore declines as the disc accretes, and a power-law decline in $\Mdot(t)$ is inevitable in any viscous model with approximately constant $\alpha$. By contrast, the geometric decline of $\Mdot(t)$ in the wind-driven model arises from the choice of disc wind model. We follow the prescription of \citet{tabone22b}, which corresponds to the ``$\Sigma$-dependent $\alpha_{\mathrm {DW}}$'' model in \citet{tabone22a}. In this model the accretion efficiency $\alpha_{\mathrm {DW}}$ increases as $\Sigma$ declines. As the accretion rate $\Mdot \propto \alpha_{\mathrm {DW}} \Sigma$, physically this corresponds to a wind-driven accretion rate that has only a weak dependence on $\Sigma$\footnote{Indeed, the wind-driven accretion rate is completely independent of $\Sigma$ in the limiting case $\omega = 1$.}. 

In strict terms our analysis therefore tests how strongly the accretion efficiency ($\alpha$) depends on disc surface density, rather than directly probing the mechanism driving the accretion, with the viscous $\alpha$ assumed to be independent of $\Sigma$. Simulations of ideal MHD turbulence in fully-ionized, strongly-magnetised discs find that $\alpha \propto \beta^{-1/2}$, where the dimensionless plasma $\beta$ parameter is defined as the ratio of gas to magnetic pressure \citep{salvesen16}. However, this scaling is not reproduced in simulations with zero net magnetic flux, and in conditions typical of protoplanetary discs no strong scaling with $\Sigma$ is observed \cite[see, e.g., discussion in ][]{lesur_pp7}. In reality $\alpha$ varies with a number of different parameters (most notably the poloidal magnetic field strength), but as long as there is no strong dependence on $\Sigma$, a power-law decline in $\Mdot(t)$ is the natural outcome of viscous accretion in protoplanetary discs.

On the other hand, the accretion {\it rate} $\Mdot$ being largely independent of $\Sigma$ is seen in a range of wind-driven disc evolution models \citep[e.g.,][]{armitage13,suzuki16}. Numerical simulations find that the rate of wind-driven accretion depends primarily on the magnetic field strength, with only a weak dependence on $\Sigma$ \citep[e.g.,][]{bai13}, though the transport of magnetic flux remains a significant uncertainty \citep[see][and references therein]{lesur_pp7}. Therefore, although the difference between the models formally arises from the assumed scalings of $\alpha$ with surface density, we conclude that the qualitative difference in the accretion rate distributions from our two disc evolution models is a robust physical prediction.

The key assumption in our method is that the observed accretion rates are representative of the underlying distribution. For a specific disc model this implies uniform sampling in time (as in Fig.\,\ref{fig:fake_data}), but in a population of discs with a spread in lifetimes this does not translate directly to stellar age. Moreover, clusters of young stars have spreads in ages, so the validity of our assumption is difficult to quantify. In practice we require that the observed accretion rates are representative of the population, and that that population is not observed at a special time in the discs' evolution. For a sample drawn from many different star-forming regions (like that of \citealt{manara_pp7}) this seems reasonable, but this does represent a potential systematic uncertainty in our approach. The treatment of ``non-accreting'' Class III sources is also an issue, as our models do not consider stars which no longer have discs. At present the distribution of accretion rates does not change significantly when the upper limits are included (see Fig.\,\ref{fig:pp7_histogram}), but future analyses may need to distinguish between low $\Mdot$ discs, and disc-less stars which are not accreting at all. 

An additional concern is how robust our statistical results are against changes in the model parameters. The comparison in Section \ref{sec:ks_tests_scatter} assumes a canonical set of median parameters for both models. These are motivated by previous demographic modelling \citep[e.g.,][]{aa09,tabone22b}, but the statistical comparison is somewhat sensitive to the choice of median parameters. In particular, the models become harder to distinguish (requiring sample sizes 2--3 times larger) if the initial accretion rate in the wind-driven model is reduced by 0.5--1.0\,dex. However, such a choice of parameters is disfavoured by previous studies (as it requires either low disc masses or long disc lifetimes), and it also means that the model fails to reproduce the highest observed accretion rates (see Fig.\,\ref{fig:pp7_histogram}). As long we require that our input models are consistent with other demographic indicators, then a few hundred sources is sufficient to distinguish between them at high confidence.

Alternatively, we could in principle adopt a data-driven approach and invert the problem, using the observed accretion rates to specify the distribution of model input parameters. This is not possible with the existing data, but a larger sample would yield distributions of input parameters for both models. The question then becomes whether or not the derived parameters are consistent with other observations (such as disc lifetimes and/or disc masses). Our method is therefore not strictly independent of these other demographic indicators. For a given model set the distribution of the accretion rates can be used as a stand-alone diagnostic, but our approach relies on other observables to define a ``reasonable'' range of input parameters for the evolutionary models.

\subsection{Implications for disc evolution}
Our results are broadly consistent with existing demographic modelling, though with some interesting differences. The main qualitative difference between our approach and previous studies \citep[summarized in][]{manara_pp7} is that we consider only the accretion rates, and do not draw any inferences from other evolutionary indicators (such as stellar ages, disc masses, or disc sizes). Our method essentially marginalises over time, and as a result it is largely insensitive to the absolute time-scales (and also therefore the the magnitude of $\alpha$); the benefit of our approach lies in its statistical power, and in the much smaller systematic uncertainties in the observables. Population synthesis modelling has previously shown that viscous/photoevaporation models are broadly consistent with observed accretion rates and disc fractions \citep[e.g.,][]{aa09,owen10,jones12,mulders17,picogna19}, and our canonical parameters are based on the conclusions of these studies. However, \citet{lodato17} found that viscous models can only reproduce the observed relationship between accretion rate and disc mass if the viscous time-scale is relatively long ($t_{\nu} \simeq 0.3$--$1\times10^6$\,yr). This is a factor 5--10 larger than in our canonical model, and contradicts our earlier assumption that $t \gg t_{\nu}$. A long viscous time-scale does not invalidate our approach, but would add one extra parameter to the model (as $p(\Mdot)$ is no longer independent of $\Mdot_0$). However, part of this apparent discrepancy is simply due to the choice of the viscosity index $\gamma$\footnote{Changing from $\gamma = 3/2$ (preferred by \citealt{lodato17}) to $\gamma = 1$ (as in our canonical model) reduces $t_{\nu}$ by a factor of 4.}, and \citet{somigliana20} showed that including photoevaporation also weakens the requirement for $t_{\nu}$ to be long (by adding scatter to the relation; see also \citealt{sellek20}). We therefore do not consider our results to be in significant disagreement with \citet{lodato17}. We also note that these two analyses used independent demographic indicators, so somewhat different results are not unexpected. If the discrepancy is real, then it may reflect systematic uncertainties in some of the observables (e.g., in the disc masses).

Demographic models of wind-driven accretion are a recent development, so are less well-studied than viscous models. \citet{tabone22b} showed that wind-driven accretion can reproduce the observed decline in disc fraction with age, as well as rapid disc clearing at the end of the disc lifetime, while \citet{trapman22} showed that the evolution of disc sizes is also consistent with wind-driven accretion. Our canonical model is based on these studies, and our results are broadly similar. We note, however, that reproducing the upper end of the observed accretion rate distribution (at $\gtrsim 10^{-7}$\,\Msunyr) requires $\Mdot_0$ to be at least an order of magnitude larger than in the canonical model of \citet{tabone22b}. This in turn requires either that wind-driven accretion is extremely efficient, or that there is a very large spread of initial conditions in wind-driven discs (spanning two orders of magnitude in the initial accretion rate). Recently, \citet{long22} and \citet{zagaria22} both compared observations of disc sizes with both viscous and wind-driven models, and both found that wind-driven accretion is weakly favoured. This suggests that a combined statistical study of both disc sizes and accretion rates may provide interesting additional insights. 

The only parameter which is significantly constrained by our analysis is the cut-off accretion rate in the viscous model: we see from Fig.\,\ref{fig:mdot_c} that $\Mdot_{\mathrm c} \lesssim 10^{-9}$\,\Msunyr\ (see also Section \ref{sec:stats}). Physically this represents the mass-loss rate due to photoevaporation. Our results are agnostic as to the mechanism driving disc photoevaporation, but place a strict upper limit on the mass-loss rate. This is consistent with the results of previous demographic studies (e.g., \citealt{alexander12,somigliana20,manzo-martinez20}; see also \citealt{rda_pp6}), but our statistical analysis provides a much stronger limit on the mass-loss rate than has previously been possible. However, low rates of photoevaporative mass-loss are inconsistent with the predictions of X-ray photoevaporation models, which typically find mass-loss rates $\sim$\,$10^{-8}$\,\Msunyr\ \citep[e.g.,][]{owen10,picogna19}. This conclusion is readily understood: the median accretion rate in the observed population is $\Mdot \sim 10^{-9}$\,\Msunyr, with a large number of observed discs accreting at lower rates, and any model which shuts off disc accretion above the median observed $\Mdot$ is clearly ruled out\footnote{Recently \citet{atom23} found evidence for an even lower cut-off in the $\Mdot$ distribution, at $\simeq 10^{-10}$\,\Msunyr.}. If accretion at AU radii is not viscous then this discrepancy could be resolved, but otherwise the data point unambiguously towards photoevaporation rates $\lesssim 10^{-9}$\,\Msunyr. 

In reality it seems likely that all three processes operate at different locations and times during protoplanetary disc evolution. Current understanding points towards turbulent transport being efficient at small radii, accretion being primarily wind-driven elsewhere in the disc, and photoevaporation driving final disc dispersal \citep{lesur_pp7,pascucci_pp7}. Our models are idealised, and future work should consider how these different processes interact, and how we can potentially diagnose their effects in ``hybrid'' models. Future observations at low $\Mdot$ -- in particular a homogeneous sample of upper limits in ``non-accreting'' Class III discs -- will be critical to applying our new method more widely, as the statistical power of our current analysis is limited by the lack of useful data points at $\Mdot$\,$\lesssim$\,$10^{-10}$\,\Msunyr. A larger sample of higher accretion rates $\gtrsim$\,$10^{-7}$\,\Msunyr\ would also provide useful constraints on the physics of wind-driven accretion.


\section{Summary}
In this paper we have proposed that the distribution of observed disc accretion rates can be used as a stand-alone diagnostic of protoplanetary disc evolution. We have shown that the differing transport processes in turbulent (``viscous'') transport of angular momentum and wind-driven accretion result in fundamentally different distributions of accretion rates. Our Monte Carlo analysis shows that these distributions are distinguishable at high confidence with relatively small sample sizes ($N \gtrsim 300$), even for pessimistic assumptions about the scatter in the model parameters. This approach assumes that the observed accretion rates are representative of the underlying distribution, but relies only on a single, well-measured observable (the stellar accretion rate). It therefore offers significant advantages over other demographic methods, which rely on disc properties (notably stellar ages and disc masses) which are subject to large systematic uncertainties. 

We then applied our method to observations, using the compilation of accretion rates from \citet{manara_pp7}. We find that current data do not provide a large enough sample to distinguish between the models clearly, but a modest increase in the number of observed accretion rates, as well as a homogeneous sample of upper limits in non-accreting sources, would increase the statistical power of this sample significantly. In the case of viscous accretion, the large number of discs with low observed accretion rates limits the rate of disc photoevaporation to be $\lesssim 10^{-9}$\,\Msunyr. Accretion rates therefore offer a direct observational test of disc evolution, and uniform, homogeneous surveys of accretion rates can provide a clear answer to the question of how protoplanetary discs accrete. 


\section*{Acknowledgements}
We thank the anonymous referee for an insightful review.
RA and GR acknowledge funding from the Science \& Technology Facilities Council (STFC) through Consolidated Grant ST/W000857/1. 
GR acknowledges support from the Netherlands Organisation for Scientific Research (NWO, program number 016.Veni.192.233) and from an STFC Ernest Rutherford Fellowship (grant number ST/T003855/1). 
PJA acknowledges support from NASA TCAN award 80NSSC19K0639. 
GJH is supported by General Grant 12173003 from the National Natural Science Foundation of China. 
BT is a Laureate of the Paris Region fellowship program, which is supported by the Ile-de-France Region and has
received funding under Marie Sk\l odowska-Curie grant agreement No.\,945298.
This project has received funding from the European Research Council (ERC) under the European Union’s Horizon Europe Research \& Innovation Programme under grant agreements No.\,101039651 (DiscEvol) and No.\,101039452 (WANDA). Views and opinions expressed are however those of the author(s) only and do not necessarily reflect those of the European Union or the European Research Council Executive Agency. Neither the European Union nor the granting authority can be held responsible for them.

\section*{Data Availability}
The observational data used in this paper are from the compilation of \citet{manara_pp7}, and are publicly available at \url{http://ppvii.org/chapter/15/}. The randomly-generated data from our statistical analysis will be shared on reasonable request to the corresponding author.


\bibliographystyle{mnras}


\bsp	
\label{lastpage}
\end{document}